# scientific reports

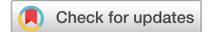

OPEN

# Solar UV-B/A radiation is highly effective in inactivating SARS-CoV-2

Fabrizio Nicastro[1]✉, Giorgia Sironi[2], Elio Antonello[2], Andrea Bianco[2], Mara Biasin[3], John R. Brucato[4], Ilaria Ermolli[1], Giovanni Pareschi[2], Marta Salvati[5], Paolo Tozzi[4], Daria Trabattoni[3] & Mario Clerici[6]

Solar UV-C photons do not reach Earth's surface, but are known to be endowed with germicidal properties that are also effective on viruses. The effect of softer UV-B and UV-A photons, which copiously reach the Earth's surface, on viruses are instead little studied, particularly on single-stranded RNA viruses. Here we combine our measurements of the action spectrum of Covid-19 in response to UV light, Solar irradiation measurements on Earth during the SARS-CoV-2 pandemics, worldwide recorded Covid-19 mortality data and our "Solar-Pump" diffusive model of epidemics to show that (a) UV-B/A photons have a powerful virucidal effect on the single-stranded RNA virus Covid-19 and that (b) the Solar radiation that reaches temperate regions of the Earth at noon during summers, is sufficient to inactivate 63% of virions in open-space concentrations ($1.5 \times 10^3$ $TCID_{50}$/mL, higher than typical aerosol) in less than 2 min. We conclude that the characteristic seasonality imprint displayed world-wide by the SARS-Cov-2 mortality time-series throughout the diffusion of the outbreak (with temperate regions showing clear seasonal trends and equatorial regions suffering, on average, a systematically lower mortality), might have been efficiently set by the different intensity of UV-B/A Solar radiation hitting different Earth's locations at different times of the year. Our results suggest that Solar UV-B/A play an important role in planning strategies of confinement of the epidemics, which should be worked out and set up during spring/summer months and fully implemented during low-solar-irradiation periods.

It is well known that 200–290 nm ultraviolet photons (hereinafter UV-C radiation) are absorbed by both the nucleic acid (RNA and DNA) and proteins and are endowed with germicidal properties that are also effective on viruses because they induce irreversible modifications that inhibit the virus replication[1–9] (see "Methods"). Fortunately, Solar UV-C photons of this wavelength are filtered out by the Ozone layer of the upper Atmosphere, at around 35 km[10]. Softer UV photons from the Sun, however, do reach the Earth's surface. In particular, about 10% of the photons emitted by the Sun in the range 290–315 nm (UV-B) and as much as 95% of those in the range 315–400 nm (UV-A), reach the Earth. The effect of these photons on Single- and Double-Stranded RNA/DNA viruses[10–13] and the possible role they play on the seasonality of epidemics[14–25], are nevertheless little studied and highly debated in alternative or complementarity to other environmental causes[23,26–39]. Notably though, the effects of both direct and indirect radiation from the Sun needs to be considered in order to completely explain the effects of UV radiations in life processes (with e.g. the UV virucidal effect enhanced in combination with the concomitant process of water droplets depletion because of solar heat and/or reduced in combination with high-humidity because of the larger optical depth).

Herein, and in a companion paper that presents our laboratory measurements[40], we show that the action spectrum of Covid-19 in response to photons with λ = 366–405 nm (UV-A) is surprisingly efficient. As a consequence, UV-B/A integrated direct light from the Sun on the Earth have a strong disinfection power on the Severe Acute Respiratory Syndrome CoV-2 (SARS-CoV-2), inactivating 63% of the virus in $1.5 \times 10^3$ $TCID_{50}$/mL concentrations (higher than typical aerosol concentrations) in about 1–2 min at noon during north/south summers or at the Equator. We also present a number of concurring circumstantial evidence suggesting that the evolution and strength of the recent SARS-Cov-2 pandemics[41,42], has been (and is being) modulated by the intensity of UV-B and UV-A Solar radiation hitting different regions of Earth during the diffusion of the outbreak between January 2020 and April 2021. Finally, we show that, once the laboratory-calibrated effect of the "Solar-Pump" is included, our SIRD-like models of the epidemics[20] are able to qualitatively reproduce the observed time-resolved mortality of the SARS-CoV-2 epidemics in different regions of the Globe. Our findings could help in designing the social behaviors to be adopted depending on season and environmental conditions.

[1]Italian National Institute for Astrophysics (INAF)-Rome Astronomical Observatory, Rome, Italy. [2]Italian National Institute for Astrophysics (INAF)-Brera Astronomical Observatory, Merate, Milan, Italy. [3]Department of Biomedical and Clinical Sciences L. Sacco, University of Milano, Milan, Italy. [4]Italian National Institute for Astrophysics (INAF)-Arcetri Astrophysical Observatory, Florence, Italy. [5]Regional Agency for Environmental Protection of Lombardia (ARPA Lombardia), Milan, Italy. [6]Department of Pathophysiology and Transplantation, University of Milano, Don C. Gnocchi Foundation, IRCCS, Milan, Italy. ✉email: fabrizio.nicastro@inaf.it

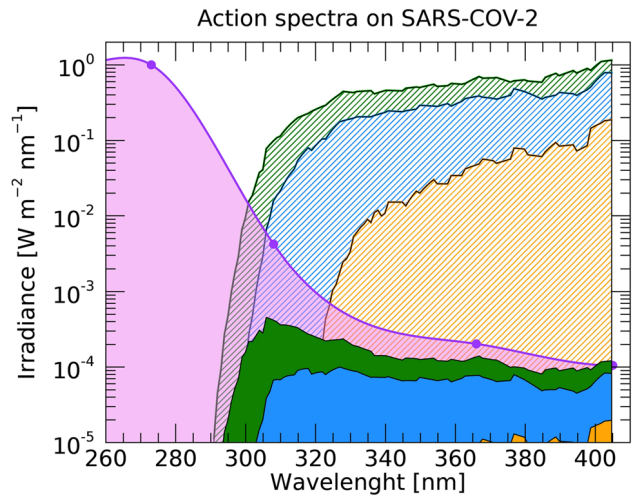

**Figure 1.** 254 nm-normalized action spectrum of Covid-19 to D63 UV doses (pink curve and circles), solar spectra at zenith angles of 0 (green curve), 60 (blue curve) and 80 (orange curve) degrees, and integrals of their product, i.e. Solar UV Inactivation Powers (SUVIPs, green, blue and orange shaded area). Clearly most of the SUVIP is in the UV-A regime (315–400 nm), and is a strong function of the zenith angle (or Earth latitude).

## Results

Coronavirus infections were shown to be influenced by seasonal factors including UV-index, temperature and humidity (e.g.[43,44]). However, whether these factors, either independently or combined together, are able to directly modulate the geographical pandemics development of the ongoing SARS-CoV-2 infection is still debated (e.g.[45,46], but see also[22,33–39]). Yet, the available data on SARS-CoV-2 contagions show indisputably that the ongoing pandemic has a lower impact during summer time and in countries with a high solar illumination (e.g.[15,19,20,22,24,47–49]). This suggests that UV-B/A Solar photons may play a role in the evolution of this pandemic, perhaps in combination with more subtle, and related, indirect factors such as temperature and humidity.

The optical-ultraviolet solar spectrum peaks at a wavelength of about 500 nm, but extends down to much more energetic photons of about 200 nm. Due to photo-absorption by the upper Ozone layer of the Earth's atmosphere, however, only photons with wavelengths longer than ~290 nm (those not sufficiently energetic to efficiently photo-excite an $O_3$ molecule and dissociate it into an $O_2 + O$), reach the Earth's surface[50]. For a given exposure, the daily fluence (flux integrated over a daytime) on Earth of this UV radiation depends critically on Earth location (latitude) and time of the year and results in the potential of Solar UV-B/A photons to inactivate a virus in air (aerosol) or surfaces.

To evaluate this potential on the onset, diffusion and strength of the recent SARS-CoV-2 pandemics, in our analysis we use the Covid-19 action spectrum (i.e. the ratio of the virucidal dose of UV radiation at a given wavelength to the lethal dose at 254 nm see "Methods") recently compiled by us in the laboratories of the University of Milan ([40]; see "Methods"), together with UV-B and –A solar radiation measurement estimates on the Earth as a function of Earth's latitude and day of the year ([51]; see "Methods").

Results are surprisingly encouraging. Figure 1 (and Table 1) shows the 254-nm-normalized action spectrum of Covid-19 to $D_{63}$ UV doses (the amount of UV photons needed to inactivate 63% of the virions), in the 260–405 nm band (UV-C through UV-A light), together with state-of-the-art models of the Solar spectra on the Earth in the same wavelength range[52] at three representative zenith angles and the integrated Solar UV Inactivation Power (SUVIP), i.e. the actual Solar radiation power available to inactivate the virus (see Figure's caption and "Methods" for details).

The action spectrum of Covid-19 in response to photons with λ = 366–405 nm (in the UV-A) is at least 2 orders of magnitude more efficient than that of direct DNA-damage (see e.g. https://www.temis.nl/uvradiation/product/action.html) or that obtained by extrapolating to longer wavelengths the action spectrum of[12] (derived for a compilation of a wide variety of viruses). As a consequence, "instantaneous" Solar lethal-times $\tau_{63}$, i.e. the time needed for the Sun to deliver a lethal dose $D_{63}$ at a given time of the day and location on the Earth, are relatively short. These, can be easily derived by simply dividing the normalizing measured Covid-19 UV-C lethal doses $D_{63}(254) = 3.5$ W m$^{-2}$ (Table 1) by actual SUVIPs, as recorded at a given location on the Earth and a given time of the year (see "Methods" for details).

| Central λ (in nm) | $D_{63}$ (in j m$^{-2}$) | $D_{90}$ (in j m$^{-2}$) | 63% action-spectrum |
|---|---|---|---|
| 254 (UV-C) | 3.5 | 5.0 | 1 |
| 278 (UV-C) | 3.5 | 5.0 | 1 |
| 308 (UV-B) | 524 | 746 | $6.6 \times 10^{-3}$ |
| 366 (UV-A) | 16,488 | 23,477 | $2.1 \times 10^{-4}$ |
| 405 (near UV-A) | 22,434 | 31,942 | $1.6 \times 10^{-4}$ |

**Table 1.** UV doses needed to inactivate 63% and 90% of Covid-19 virus in $1.5 \times 10^3$ TCID$_{50}$/mL solutions and derived 63% action spectrum of Covid-19.

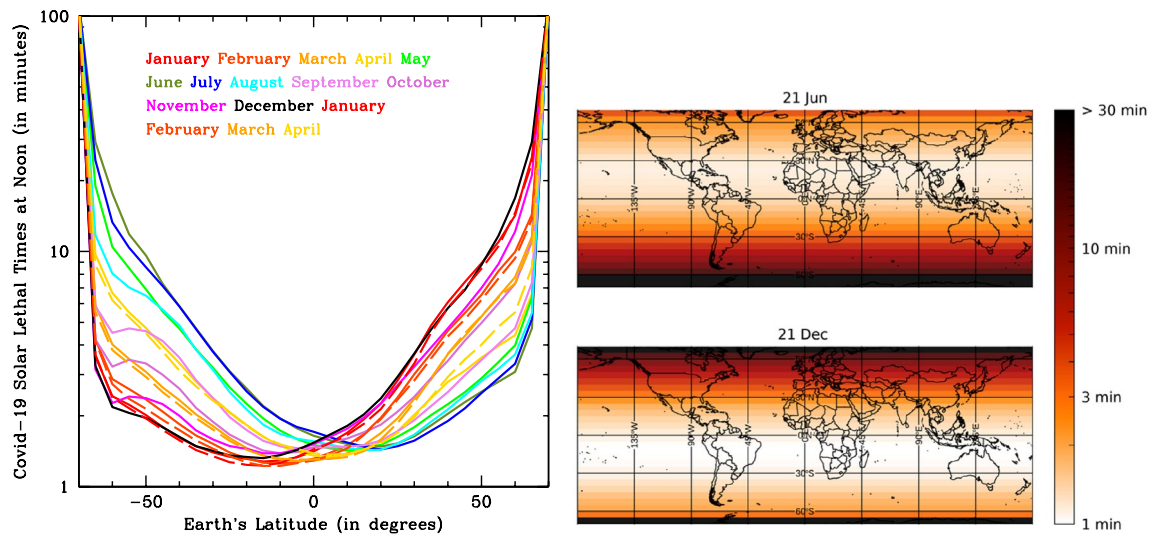

**Figure 2.** (**a**) Covid-19, monthly-averaged, lethal-times at noon in latitude bins of 5 degrees, from -70 to +70 degrees, for year 2020 and the first four months of 2021 (left panel). (**b**) Two world maps with a superimposed $\tau_{63}$ intensity-gradient image, at summer (top panel) and winter (bottom panel) solstices (right panel).

Curves of Fig. 2a shows Covid-19 monthly-averaged lethal-times at noon in latitude bins of 5 degrees, from −70 to +70, for year 2020 and the first 4 months of 2021 (see "Methods" for details). The weak asymmetry between the South and North hemispheres is mostly due to the smaller distance of the Earth from the Sun during south summers compared to north summers (7% more Solar flux at perihelion than aphelion). Less than 2 min are needed at noon, all year round, to inactivate 63% of Covid-19 in aerosol or on surfaces in equatorial regions (|α|< 20 degrees). This is also the case for temperate-region locations (20 <|α|< 60) during boreal and austral summers (i.e. June–August in the North and December–February in the South), but then lethal times in these regions start increasing steadily by up to factors of 20 from falls through winters, to then decrease again in Spring. Figure 2b shows the same thing through imaging: two world maps with a superimposed $\tau_{63}$ intensity-gradient image, at summer (top panel) and winter (bottom panel) solstices.

To investigate the effect of this radiation on the diffusion and strength of the SARS-CoV-2 pandemics, we collected Covid-19 data from the GitHub repository provided by the Coronavirus Resource Center of the John Hopkins University (see Figure's caption and "Methods" for details).

Figure 3 shows that the absolute maximum number of daily SARS-CoV-2 deaths per million inhabitants (mortality) reached in each country of our Sample, anti-correlates with the Solar-UV inactivation index Ξ on the Earth 3 weeks prior to the mortality peak (see Figure's caption and "Methods" for details).

We did not attempt to correct the data for social-distancing measures (adopted differently and at different times of the years in different regions of the world) or for tropical daily cloud-coverage (where present systematically, like in equatorial Countries during rain versus dry seasons), which both certainly contribute significantly to the scatter visible in the data (main plot) and, indeed, greatly reduced when these are grouped in either latitude or Solar Inactivation Index bins (left and right onsets, respectively). Nonetheless, the correlation in Fig. 3 is statistically highly significant and extends more than one order of magnitude in Ξ and over four orders of magnitude in mortality. Simple linear-regression tests of the data-points (Pearson) and their rankings (Spearman), yield null (i.e. chance correlation) probabilities of $p = 6.4 \times 10^{-12}$ and $p = 2.0 \times 10^{-12}$, respectively, corresponding to Gaussian-equivalent significances of 6.8σ and 7σ.

In Fig. 4 we show the observed mean mortality time series of the three large N, T and S groups of Countries of our sample (see Figure's caption and "Methods" for details). Clearly, not only a pronounced modulation with seasons on Earth is observed in the N group (and less evidently in the S group), but more importantly, the average mortality in the N group during north autumns-winters is up to 8 times higher than the average mortalities in the T and S groups (solid curves). Correspondingly, Covid-19 lethal-times are up to 5 times longer (dotted curves). This is virtually impossible to explain in terms of differences in the efficiency of lockdowns or social-distancing measures, unless one speculates that the large number of countries populating the T and S groups have concordantly and simultaneously adopted much more strict measures than those adopted by all countries in the N group. An external common cause must be invoked, which broadly follow the pace of seasons on the Earth.

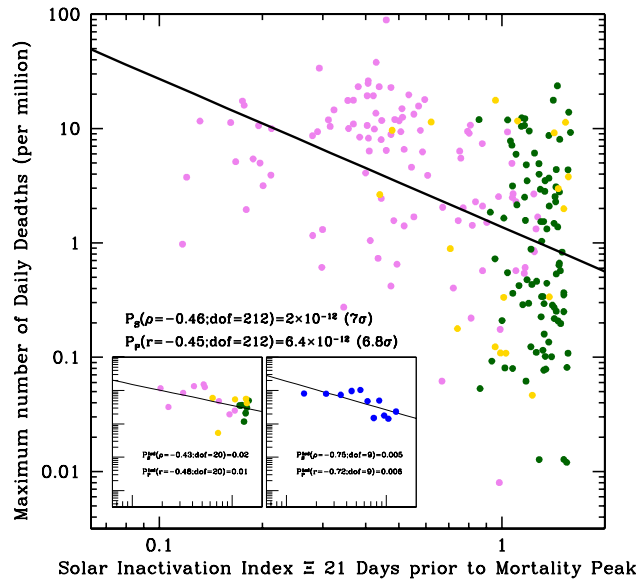

**Figure 3.** Absolute maximum number of daily SARS-CoV-2 deaths per million inhabitants for the 214 Countries (or territories) of our sample, versus the Solar UV Inactivation index Ξ on the Earth 3 weeks prior to the mortality peak. The Solar UV Inactivation Index Ξ is defined as the ratio of the SUVIP administered in 2 min to the Covid-19 lethal dose D63. (see "Methods"). Pink, green and yellow circles are Countries of the north hemisphere ($\alpha > 20$ degrees: 103 Countries; N group), the equatorial strip ($|\alpha| \leq 20$ degrees: 93 Countries; T group), and the south hemisphere ($\alpha < -20$ degrees: 18 Countries; S group), respectively. Grouping the data in either 5-degree-wide latitude bins (left onset) or 0.1-wide Ξ bins (right onset) greatly reduces the scatter present in the main plot, but also the degrees of freedom of the relations, thus yielding to higher (but still low) probabilities of chance correlation.

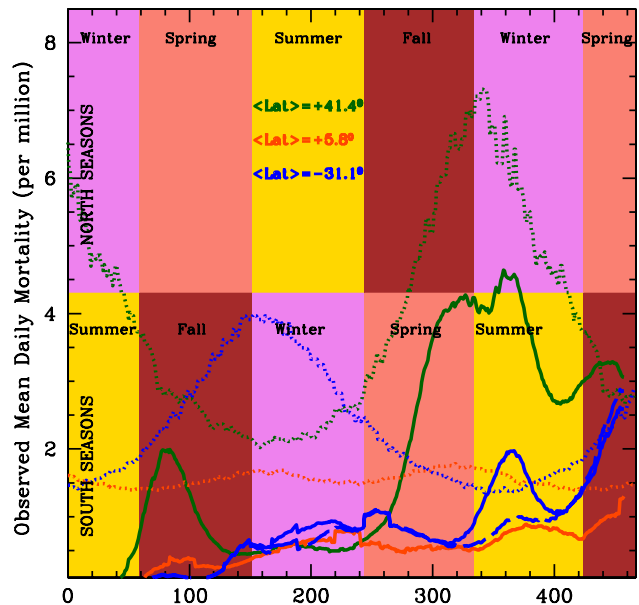

**Figure 4.** Observed mean mortality time series (see "Methods" for details) of the three large groups of Countries at latitudes $\alpha > 20$ degrees (N group: green solid curve), $-20 \leq \alpha \leq 20$ degrees (T group: orange solid curve), and $\alpha < -20$ degrees (S group: blue solid curve). Dotted curves are Covid-19 Lethal-Time solar light-curves at the three average latitudes $\alpha = 41.4$ degrees (N; green), $\alpha = 5.8$ degrees (T; orange) and $\alpha = -31.1$ degrees (S; blue). The long-dashed blue curve of this figure shows the mean mortality curve of all countries of the S group after excluding South Africa and Eswatini.

It is also important to note that the rapid and short-lasting (about 30 days) factor of 2 increase in mortality in the S group toward the end of 2020 (beginning of south summer), followed by a sharp decline in February 2021, is largely dominated by South-Africa and Eswatini (a small landlocked country embedded within South Africa), with peaks of 9 and 11 deaths per million inhabitants around mid January 2021. This is where and when the SARS-CoV-2 B.1.351 variant, associated with higher viral load and greater transmissibility potentials (https://www.paho.org/en/documents/epidemiological-update-occurrence-variants-sars-cov-2-americas-20-january-2021), has first developed. The long-dashed blue curve of Fig. 4 shows the mean mortality curve of all countries of the S group after excluding South Africa and Eswatini.

## Discussion

As expected by our model, in South and Tropical Countries the onset of the outbreak was homogeneously distributed from the end of March through early May (when typical UV-B/A instantaneous virus-inactivation exposures at noon are between 1.5 and 3 min: blue and orange dotted curves of Fig. 4, respectively), and the average daily mortality rate (excluding South Africa and Eswatini: blue dashed curve of Fig. 4) remained limited to a maximum of 1 per million per day throughout all 2020 and the first monitored part of 2021. The average daily mortality curve in countries of the S group (blue dashed curve in Fig. 4) smoothly modulates between the end of south-autumn throughout the beginning of south-summer (with a maximum towards the end of south winter 2020), and then starts climbing more decisively toward the end of south-summer 2021, when also lethal times increase. The average daily mortality curve of equatorial countries, instead (solid orange curve of Fig. 4), stays roughly constant at a low level of ~ 0.5–1 deaths per million throughout all the monitored period.

On the contrary the onset of the epidemics in countries of the N group anticipates that of countries in the T and S groups by at least one month, and the average daily mortality rate curve of these countries follows a typical seasonal trend (solid green curve in Fig. 4). It rises quickly during March 2020 but is efficiently suppressed by the strict lockdown measures implemented by virtually all countries in the N group during the first wave of the epidemics. Such suppression generates a peak around mid-April 2020 at a relatively moderate level of 2 deaths per million (more than twice larger than the corresponding rates in the T and S groups), followed by a sharp decline between April and May 2020. This decline takes the average daily mortality rate in countries of the N group down to a minimum level of about 0.5 deaths per million by the end of May, beginning of June 2020, and is followed by a long steady phase that lasts throughout all summer 2020, despite most of the social-distancing measures in the majority of countries of the N group had been greatly relieved already by the beginning of June. During this long, steady low-mortality phase, average solar lethal-times at noon in countries of the N group are ~ 2.5 min and day-light coverage ~ 10–12 h (see also onsets of Fig. S1 of Supplementary Information: SI). Finally, in October 2020 the daily mortality rate starts increasing again and by February 2021 reaches maximum levels 7–8 times those recorded in summer (despite in the majority of the Countries of this group social-distancing measures—generally milder than those adopted during the first wave in winter 2020—had been already enforced since late fall 2020), while correspondingly the average solar lethal-time at noon in this group of Countries increases by up to a factor of 3.5 and day-light coverage shrinks by a factor of about 2 (see onsets in Fig. S1 of SI).

The segregation by Earth's latitude of the curves of growth of Fig. 4 can be well reproduced (both qualitatively and quantitatively) by simulating the effect of solar virus inactivation with Monte-Carlo runs of our "Solar-Pump" diffusive model[20] (see "Methods" for details).

Figure S1 of SI shows typical yearly (main panel) and infra-day (onsets) $R_t$ modulations of an intrinsic reproduction number $R_0 = 3$, imprinted by the solar mechanism (see details in Figure's caption and "Methods").

In Figure S2 of SI we show the results of our Monte Carlo simulations, with (solid curves) and without (dashed curves) the effect of the Solar-Pump (see "Methods" for details). The X and Y scales in Fig. S2 of SI and Fig. 4 are exactly the same, for an easy comparison between actual data and simulations.

Clearly, only when an external mechanism acting with the pace and latitude-dependent intensity of seasons on Earth, is at work (solid curves) the observed average mortality time-series in the three groups of countries can be, at least qualitatively, reproduced (compare solid curves of Fig. S2 in SI with solid curves of Fig. 4). When the Solar-Pump mechanism is not active (dashed curves), the average daily mortality time-series in the N, T and S groups reaches peaks of 57, 93 and 66 deaths per million, respectively, far higher than observed, and only the N group, in these simulations, show a second, much lower (only ~ 1.5 daily death per million) peak in November 2020. Such high maxima could be suppressed to the observed levels by ad-hoc efficiencies of the, country-by-country, adopted social-distancing measures, but it would require extreme fine-tuning of the starting/stopping and halving/doubling times of measures' implementation, loosening and duration in the three groups.

Taken together, the curves of growth of Fig. 4, and their comparison with those of Fig. S2 in SI, suggest that the SARS-CoV-2 mortality is efficiently suppressed in areas where typical UV-B/A virus-inactivation exposures at noon are shorter than about 2–3 min (i.e. Solar UV Inactivation index Ξ higher than 0.7–1).

This value should be compared to the stability of SARS-CoV-2 on aerosol and surfaces. A recent study[53] shows that SARS-CoV-2 can survive in aerosols for longer than three hours and is even more stable on metallic and polymeric surfaces, where it can stay for up to several tens of hours and is quickly inactivated by solar light[54]. These results, combined with another recent research showing that the lifetime of speech droplets with relatively large size (> few micrometers) can be tens of minutes[55], suggest that SARS-CoV-2 may efficiently spread through aerosols (either directly, through inhalation, and/or indirectly, through evaporation of contaminated surfaces).

Seasonal air temperature oscillations on temperate regions of the Earth, as well as large day/night thermal excursion in dry and low-latitude regions, are due to the different amount of Solar radiation hitting different locations on the Earth's surface at different times of the year. The two quantities are therefore obviously broadly correlated. However, at Earth's temperature heat itself cannot be the direct cause of virus inactivation in summers or at the equator in open-spaces, since Covid-19 is known to survive to non-Earth-like temperatures of ~ 60–70 C, for up to 10 min and to continuous exposure to extreme, but more common, summer-like temperatures of 37 C for longer than 6 h ([56]).

However, heat (i.e. high air temperatures), in low-humidity environments, can certainly boost the direct disinfecting power of Solar UV-B/A radiation, by speeding up the evaporation of water droplets containing virions (e.g.[31]).

Humidity, instead, is likely to play against Solar UV-B/A inactivation power, by both hampering efficient water droplet evaporation and by increasing the optical depth to disinfecting radiation. This has indeed been observed (e.g.[44,45]), particularly in equatorial regions where a year-round risk of coronavirus infection is present ([44]; Fig. 4), albeit with consistently lower mortality, due probably to the combined effect of the high and almost constant amount of UV-B/A radiation administered by the Sun over time, and the increased optical depth due to high humidity.

The high efficiency of Solar UV inactivation power against Covid-19, had already been suggested by previous works, based on either general models on coronaviruses (e.g.[13]), observed anti-correlation between sunlight doses and SARS-CoV-2 positive cases (e.g.[19]) and/or actual experiments with simulated sunlight (e.g.[54]). Particularly, the comparison between predictions ([13]) and actual measurements of Covid-19 exposure to artificial Solar radiation ([54]), suggested the importance of UV-A photons ([16,17]). Ratnesar-Schumate and collaborators ([54]) found that 6.8 min of continuous exposure of dried saliva-like concentrations of Covid-19 to simulated sunlight in summer at a latitude of 40 degrees were sufficient to inactivate 90% of virions. Our experiments use almost monochromatic UV light on even higher, but non-dried, concentrations ($1.5 \times 10^3$ TCID$_{50}$/mL) of Covid-19 and find 90% lethal times in summer at similar latitudes a factor of about 2 shorter (Fig. 4, 63% to 90%—Table 1—scaled up green dotted curve). This difference could be at least partly due to the difficulty in properly subtracting the effect of the additional optical depth introduced by the process of desiccation of the samples prior to sunlight exposure (e.g.[57]), which might have artificially increased the exposure needed to inactivate the virions in the experiment by[54].

Our results shows that Solar UV-A/B are highly efficient in inactivating SASR-Cov-2 in aerosol and surfaces and had a role in modulating the strength and diffusion of the SARS-CoV-2 epidemics, explaining the geographical and seasonal differences that characterize this disease. They also imply, if confirmed, that the UV flux received from our Sun in open areas may represent an important disinfection factor (e.g.[54]), able to significantly reduce the diffusion of the pandemics. Now that vaccines are available, but until their coverage will not assure world-wide herd immunity, our results could help in diversifying strategies of confinement of the epidemics depending on the time of the year and environmental factors, which may provide better protection with less social costs (see e.g.[41]).

## Methods

**UV lethal doses and action spectrum of Covid-19.** Details on the Covid-19 illumination experiments performed by our group at the University of Milan and the related TCID$_{50}$ inactivation measurements of Covid-19 for UV radiation, are presented in a companion paper[40]. In those experiments, an accurate estimate of the Covid-19 action spectrum to UV light was derived by delivering three different light doses at five different wavelengths (254, 278, 308, 366 and 405 nm) on extremely high virus concentrations of $6 \times 10^6$ TCID-50/mL (in DMEM), and evaluating virus replications (using the TCID$_{50}$ technique). To evaluate the UV lethal doses needed to inactivate 63% and 90% of Covid-19 ($D_{63}$ and $D_{90}$) on virus concentrations closer to typical aerosol concentrations, an additional set of measurements was then performed at 278 nm on a much lower virus solution of $1.5 \times 10^3$ TCID$_{50}$/mL, and the lethal doses at the remaining frequencies rescaled accordingly to the measured action spectrum.

In Table 1 we report the measured $D_{63}$ and $D_{90}$ doses of almost monochromatic ($\Delta\lambda = 4$ nm) UV radiation needed to inactivate 63% and 90% of Covid-19 in solutions with $1.5 \times 10^3$ TCID$_{50}$/mL virus concentrations, and the 63% Action Spectrum of Covid-19 derived by simply dividing the measured $D_{63}(254\text{ nm}) = D_{63}(278\text{ nm})$ dose by the corresponding doses of column 2.

The 63% Action Spectrum of Table 1 (column 4) is shown in Fig. 1 (pink points), together with the "spline" polynomial interpolation of the five data points (pink curve), which conservatively under-predicts the Solar power available to inactivate the virus (see below) in the relevant 308–365 nm wavelength range, with respect to the simple linear interpolation.

**Solar UV inactivation power and index.** We define the Solar UV Inactivation Power (SUVIP) as the instantaneous amount of Solar radiation available at noon at a given location and time of the year to inactivate 63% of Covid-19, i.e. the integral, over the $\lambda = 254$–400 nm wavelength range, of the product between the Solar flux at noon and the 63% Covid-19 Action Spectrum of Fig. 1 (pink points and curve).

The Solar UV Inactivation index $\Xi$ at a given location and time of the year, is defined as the ratio of the SUVIP administered in 2 min to the Covid-19 lethal dose at 254 nm $D_{63}(254)$. Thus, $\Xi = 1$ at a given location on the Earth and time of the year, means that Solar radiation at noon, is sufficient to inactivate 63% of Covid-19 virions in solutions with $1.5 \times 10^3$ TCID$_{50}$/mL concentrations in only 2 min. $\Xi$ greater/lower than unity, imply disinfection exposures shorter/longer, respectively.

**Solar radiation and microorganisms.** The mechanisms through which UV radiation from the Sun acts on viruses and, in general, microorganisms are both endogenous and exogenous and depend on the exact wavelength of the photons. For viruses, the direct endogenous mechanism is the most important, and consists in direct absorption of the UV photon by the chromophore embedded in the virus, as the nucleic acid or protein[31]. This process is known to decrease in efficiency going from UV-C to UV-B and is generally poorly active in the UV-A (though actual UV-A efficiencies on viruses have been poorly tested so far, and recent results on Covid-19 suggest that this may not always be the case), where the exogenous mechanism is effective[47,50].

UV strongly interacts with gaseous and aerosol pollutants in the air affecting the amount of UV radiation reaching the Earth surface. Nitrogen and Sulphur oxides together with Ozone strongly absorb UV light with efficiency increasing from UV-A to UV-B[50,58]. In addition, scattering of light by particles and aerosols is proportional to their concentration and highly polluted areas show again a UV reduction effect that goes in the same direction of molecular absorption.

**Solar UV inactivation index and lethal time as a function of Earth's latitude and time of the year.** The Solar UV Inactivation Powers (SUVIPs) shown for three representative zenith angles (i.e. latitude on the Earth and time of the year at noon) in Fig. 1, are obtained by multiplying models of the solar spectra at these zenith angles by the measured Covid-19 action spectrum, and integrating over the $\lambda = 254$–$400$ nm wavelength range. This could in principle be done analytically for any location on the Earth (latitude $\alpha$) and time of the year ($t$), to obtain state-of-the-art model-based SUVIPs:

$$SUVIP_{Model}(\alpha, t) = \int_{254}^{400} f_{Model}^{Sun}(\lambda, \alpha, t) A_{Covid-19}(\lambda) d\lambda \text{ (in W m}^{-2}\text{)}. \quad (1)$$

However, actual SUVIPs critically depend on the exact shape of the Solar spectra at a given location on the Earth and time of the year, in the critical 254–400 nm region. This is severely affected by a number of time-dependent factors, including the exact composition and thickness of the Ozone layer at a given location on the Earth and date. Using actually recorded Solar data during the whole pandemic period, would certainly be more accurate.

Ideally, one would like to dispose of daily ground-collected Solar spectra $f_{Obs}^{Sun}(\lambda, \alpha, t)$ at each site where pandemic data are available and for the entire duration of pandemic. These are clearly not available. However, above-atmosphere Solar and atmospheric data are collected daily by several operating satellites and the products of the analyses of these data stored in publicly accessible archives. In our analysis, we use actual Solar UV data continuously updated and made available by the Tropospheric Emission Monitoring Internet Service (TEMIS) archive. TEMIS is part of the Data User Programme of the European Space Agency (ESA), a web-based service that stores, since 2002, calibrated atmospheric data from Earth-observing satellites. TEMIS data products are provided for 259 locations around the world and interpolated on a regular latitude grid of 5 degrees, and include measurements of ozone and other constituents (e.g. $NO_2$, $CH_4$, $CO_2$, etc.), cloud coverage (only for a small number of locations), and estimates of the UV solar flux at the Earth's surface at noon. The latter, are obtained from daily-collected above-atmosphere solar spectra by exploiting state-of-the-art physical data-driven models of the atmosphere dynamics and optical depth, but are provided only as wavelength-integrated fluxes (in W m$^{-2}$) already folded with the action spectrum of Erythema (which is also provided), i.e. as Solar UV-Erythema Powers (SUVErPs):

$$SUVErP_{Temis}(\alpha, t) = \left[ \int_{254}^{400} f_{Obs}^{Sun}(\lambda, \alpha, t) A_{Eritema}(\lambda) d\lambda \right] \text{ (in W m}^{-2}\text{)}. \quad (2)$$

Unfortunately, blind de-integration of the bare Solar spectrum $f_{Obs}^{Sun}(\lambda, \alpha, t)$ from Eq. (2), is impossible (without an informed guess on its shape, it would obviously yield to infinite solutions). We therefore estimate actual SUVIPs, for each location on the Earth and each day from 22 January 2020 through 2 May 2021, by multiplying observed TEMIS SUVErPs by the ratio of the integrated Covid-19 and Erythema action spectra (both normalized at 254 nm):

$$SUVIP_{Obs}(\alpha, t) \approx SUVErP_{Temis}(\alpha, t) \times \left[ \frac{\int_{254}^{400} A_{Covid-19}(\lambda) d\lambda}{\int_{254}^{400} A_{Eritema}(\lambda) d\lambda} \right] \text{ (in W m}^{-2}\text{)}. \quad (3)$$

This is formally not correct, but it compromises between the need for actual solar data and the lack of homogeneous ground-based world-wide Solar UV spectral coverage on each day of the pandemic covered by this study.

Observed SUVIPs are then used to obtain, as a function of Earth's latitude and time of the year, both the curves of exposure (in minutes) needed to irradiate the lethal dose (i.e. lethal-times) at noon, shown in Fig. 2:

$$\tau_{63,noon}(\alpha, t) = \frac{D_{63}(254)}{SUVIP_{Obs}(\alpha, t) \times 60} \text{ (in min)}, \quad (4)$$

and the Solar UV Inactivation Index $\Xi$ 3 weeks (estimated mean duration between onset of symptoms to death: e.g.[59]) prior to the recorded maximum SARS-CoV-2 mortality, used in Fig. 3:

$$\Xi(\alpha, t_{max}) = \frac{SUVIP_{Obs}(\alpha, t_{max} - 21)}{D_{63}(254)} \times 120 = \frac{2 \, min}{\tau_{63,noon}(\alpha, t_{max})}. \quad (5)$$

In deriving the above quantities, we adopt the following conservative assumptions:

(i) we only use TEMIS *SUVErPs* collected at locations with altitude < 400 m, where UV solar fluxes are lower because of the thicker atmosphere column.
(ii) we only use TEMIS clear-sky data (the only available on a broad global scale). Cloud coverage related to local weather certainly contributes to the large scatter present along the anti-correlation shown in Fig. 3. More importantly, ignoring the systematic, season-related, effect of cloud coverage in tropical areas during rain (high-humidity) seasons, associates systematically higher-than-real Solar inactivation indexes to a significant fractions of equatorial locations on the days when mortality maxima are recorded, thus contributing to the narrow (along the X axis) and elongated (along the Y axis) distribution of the green points in the mortality-Ξ diagram of Fig. 3.
(iii) aerosols are the main responsible for local *SUVIP$_{Obs}$* absorption and scattering and may vary greatly from location to location on Earth. TEMIS data account for this effect by reducing *SUVErPs* via a flat mean attenuation factor of 0.3. Experiments made on ground showed actual doses up to a factor 2 higher than those reported by TEMIS, implying that the adopted flat 0.3 factor is highly conservative (local aerosol measurements would be incredibly interesting to study infection diffusion in polluted areas but this would require specific data per targeted area and are beyond the scope of this analysis).

**SARS-CoV-2 pandemics data.** We collect SARS-CoV-2 pandemic data from the on-line GitHub repository provided by the Coronavirus Resource Center of the John Hopkins University (CRC-JHU: https://github.com/CSSEGISandData/COVID-19). CRC-JHU global data are updated daily and cover, currently, the course of epidemics in 226 different world countries (and regions for Australia, China and Canada), by providing the daily cumulated numbers of Confirmed SARS-CoV-2 cases, Deaths and Healings.

In our study, and to produce Figs. 3 and 4, we use CRC-JHU mortality (i.e. deaths per million inhabitants) data for the 214 locations for which greater-than-zero mortality is recorded, from 22 January 2020 (recorded start of the epidemics in the Hubei region of China) through 2 May 2021.

The Y-axis value of the points of Fig. 3 is the maximum mortality recorded in each sampled location.

The continuous curves of Fig. 4, are derived by first differentiating each of the 214 cumulative mortality time series provided by CRC-JHU, then operating a non-weighted 21-day moving average on the data to smooth non-Poissonian daily fluctuations (systematics), and finally averaging out all mortality time-series of each of the three groups of locations (N, T and S).

**Modeling the Earth's latitude segregation of SARS-CoV-2 mortality with simulations.** We use our diffuse SIRD-like (Susceptible-Infected-Recovered-Deaths) "Solar-Pump" models (described in[20]) to run Monte-Carlo simulations of SARS-CoV-2 mortality time-series.

Our model considers isolated (i.e. no cross-mixing), zero-growth (i.e. death rate equal to birth rate), groups of individuals in whom an efficacious anti-viral immune response has either been elicited (Susceptible Infected Recovered: SIR) or not (Susceptible Infected Recovered Susceptible: SIRS). The inefficiency of the natural (disease) and induced (vaccines) anti-viral immune response is parameterized through population loss-of-immunity (LOI) rates $\gamma_{in}$ and $\xi_{in}$ (zero for no LOI). Other parameters of the model are the rate of contacts $\beta$ (proportional to the reproductive parameter $R_t$, with $R_{t=0} = R_0$ being the intrinsic reproductive number of the epidemic), the rate of recovery $\gamma_{out}$, the rate of deaths $\mu$, lockdown/re-opening halving/doubling times (see below), and the efficiency $\varepsilon$ of the solar-pump (see below).

For a given set of initial and/or boundary conditions, unique solutions of the model are obtained by integrating (through a simple Fortan-90 code written by us), the following set of first-order differential equations, in which the Susceptible population (S) interacts not linearly with Infected (I) at a rate $\beta$, Recovers (R) or dies (D) with rates $\gamma_{out}$ and $\mu$, respectively, get Vaccinated (V) at a rate $\xi_{out}$, and can become again susceptible (loses immunity) with rates $\gamma_{in}$ (natural) and $\xi_{in}$ (vaccines):

$$\frac{dS}{dt} = -\beta \frac{SI}{N} - \xi_{out}S + \gamma_{in}R + \xi_{in}V$$

$$\frac{dI}{dt} = \beta \frac{SI}{N} - (\gamma_{out} + \mu)I$$

$$\frac{dR}{dt} = \gamma_{out}I - \gamma_{in}R$$

$$\frac{dD}{dt} = \mu I$$

$$\frac{dV}{dt} = \xi_{out}S - \xi_{in}V$$

For each run, we assume no loss-of-immunity ($\gamma_{in} = \xi_{in} = 0$), an intrinsic reproduction number $R_0 = 3$, a starting time of the epidemic uniformly sampled from the interval 21 December 2019 ± 15 days. As generally observed, two major episodes of enforcing ("lockdown") and subsequent loosening ("reopening") of social-distancing measures are introduced in each simulation by modulating the rate of contacts β between susceptible and infected individuals, through negative (lockdowns) and positive (re-openings) exponential factors, each with starting dates and e-folding times uniformly sampled from predefined intervals. In particular, lockdowns have starting dates uniformly sampled from the 30-day wide intervals centered on 1 March 2020 and 6 November 2020, and e-folding times uniformly sampled from the 20–26 day interval. Re-openings start randomly 48–72 days and 24–36 days after the start of the first and second lockdown, respectively, with e-folding times uniformly sampled from the 123–151 day interval (virtually everywhere in the world social-distancing measures have never been completely relaxed from May 2020 to date, thus we use reopening e-folding times longer than those of "lockdowns"). A relatively low (compared to current vaccination rates in the US, Israel and part of Europe) and steady average vaccination rate of 2 per thousand susceptible individuals (with 100% immunization)—i.e. $\xi_{out} = 0.002$—is also considered in all runs, starting on 15 January 2021.

Finally, the Solar virus inactivation mechanism is introduced in the simulations by modulating the rate of contacts β between susceptible and infected individuals through the factor $(1 - \varepsilon \sin(\theta))$, where θ is the solar elevation angle (function of Earth's latitude α, time of the day ξ and day of the year $t$) and ε is the efficiency of the mechanism. The efficiency ε is evaluated at noon at a given Earth's latitude α and day of the year $t$, and is assumed to be equal to the saturated Solar UV Inactivation index Ξ, i.e. $\varepsilon(\alpha,t) = 1$ when $\Xi(\alpha,t) \geq 1$ and $\varepsilon(\alpha,t) = \Xi(\alpha,t)$ (i.e. inversely proportional to the solar lethal times in units of 2 min) otherwise. Model solutions have a time resolution of 1 day and infra-day modulations of the solar–pump efficiency are obtained by integrating daily to a sub-resolution of ~ 1.5 min.

Figure S1 in SI shows typical yearly (main panel) and infra-day (onsets) $R_t$ modulations of an intrinsic reproduction number $R_0 = 3$, imprinted by the solar mechanism (see details in Figure's caption).

Simulations are run for the same number of countries present in groups N, T and S of the actual SARS-CoV-2 data, with populations and latitudes sampled randomly from the actual population and latitude distributions within each group, and then the mortality rates of each country averaged within each group to provide the solid (models with active Solar-Pump) and dashed (models without the Solar mechanism) curves in Figure S2 of SI.

**Figures' production.** All figures are produced via scripts written by us in the scripting languages Supermongo (https://www.astro.princeton.edu/~rhl/sm/sm.html) and IDL (https://www.harrisgeospatial.com/software-technology/idl).

## Data availability

All data used in this paper are public and available at specific repositories. In particular: solar UV data are made available by the Tropospheric Emission Monitoring Internet Service (TEMIS) archive and are available at http://www.temis.nl; data of the SARS-CoV-2 pandemic are made available by the Coronavirus Resource Center of the John Hopkins University (CRC-JHUL) and downloadable from the repository https://data.humdata.org/dataset/novel-coronavirus-2019-ncov-cases; the measurements used to build the Covid-19 action spectrum in response to UV photons have been performed by our collaboration at the at the University of Milan, and data from these measurements are being published in another paper from the collaboration and are available upon request by addressing an inquire in writing to the corresponding author.

## Code availability

The code that solves numerically the system of differential equations of the models used in our analysis, has been written by the authors and is available upon request by addressing an inquire in writing to the corresponding author.

## References
1. Kowalski, W. *Ultraviolet Germicidal Irradiation Handbook: UVGI for Air and Surface Disinfection* (Springer, 2009).
2. Rauth, A. M. The physical state of viral nucleic acid and the sensitivity of viruses to ultraviolet. *Light Biophys. J.* **5**, 257–273 (1965).
3. Kesavan, J., & Sagripanti, J.L. *Disinfection of Airborne Organisms by Ultraviolet-C Radiation and Sunlight*. (Research and Technology Directorate, Edgewood Chemical Biological Center, U.S. Army, Aberdeen Proving Ground, Maryland 21010-5424 ECBC-TR-1011, 2012).
4. Chang, J. C. *et al.* UV inactivation of pathogenic and indicator microorganisms. *Appl. Environ. Microb.* **49**, 1361–1365 (1985).
5. McDevitt, J. J., Rudnick, S. N. & Radonovich, L. Aerosol susceptibility of influenza virus to UVC light. *Appl. Environ. Microbiol.* **78**, 1666–1669 (2012).
6. Welch, D. *et al.* Far-UVC light: A new tool to control the spread of airborne-mediated microbial diseases. *Sci. Rep.* **8**, 2752 (2018).
7. Walker, C. M. & Ko, G. Effect of ultraviolet germicidal irradiation on viral aerosols. *Environ. Sci. Technol.* **41**, 5460–5465 (2007).
8. Tseng, C. C. & Li, C. S. Inactivation of virus-containing aerosols by ultraviolet germicidal irradiation. *Aerosol Sci. Technol.* **39**, 1136–1142 (2005).
9. Biasin, M. *et al.* UV-C irradiation is highly effective in inactivating SARS-CoV-2 replication. *Sci. Rep.* **11**, 6260 (2021).
10. Lubin, D. & Jensen, E. H. Effects of clouds and stratospheric ozone depletion on ultraviolet radiation trends. *Nature* **377**, 710–713 (1995).
11. Furusawa, Y., Suzuki, K. & Sasaki, M. Biological and physical dosimeter for monitoring solar UV-B light. *J. Radiat. Res.* **31**, 189–206 (1990).
12. Lytle, D. C. & Sagripanti, J. L. Predicted inactivation of viruses of relevance to biodefense by solar radiation. *J. Virol.* **79**, 14244–14252 (2005).
13. Sagripanti, J. L. & Lytle, D. C. Estimated inactivation of coronaviruses by solar radiation with special reference to COVID–19. *Photochem. Photobiol.* **96**(4), 731–737 (2020).
14. Martinez, M.E. The calendar of epidemics: Seasonal cycles of infectious diseases. *PLOS Pathog.* **14**, e1007327 (2018).

15. Blum, A., Nicolaou, C., Henghes, B., & Lahav, O. On the anti-correlation between COVID-19 infection rate and natural UV light in the UK. *medRxiv* preprint. https://doi.org/10.1101/2020.11.28.20240242 (2020).
16. Luzzatto-Fegiz, P. *et al.* UVB radiation alone may not explain sunlight inactivation of SARS-CoV-2. *J. Infect. Dis.* **223**(8), 1500–1502 (2021).
17. Luzzatto-Fegiz, P. *et al.* UVA radiation could be a significant contributor to sunlight inactivation of SARS-CoV-2. *bioRxiv* preprint. https://doi.org/10.1101/2020.09.07.286666 (2020).
18. Herman, J., Biegel, B. & Huang, L. Inactivation times from 290 to 315 nm UVB in sunlight for SARS coronaviruses CoV and CoV-2 using OMI satellite data for the sunlit Earth. *Air Qual. Atmos. Health* **14**, 217–233 (2021).
19. Tang, L. *et al.* Sunlight ultraviolet radiation dose is negatively correlated with the percent positive of SARS-CoV-2 and four other common human coronaviruses in the U.S. *Sci. Tot. Environ.* **751**, 141816 (2021).
20. Nicastro, F. *et al.* Forcing seasonality of influenza-like epidemics with daily solar resonance. *iScience* **23**, 10 (2020).
21. Schuit, M. *et al.* The influence of simulated sunlight on the inactivation of influenza virus in aerosols. *J. Infect. Dis.* **221**, 372–378 (2020).
22. Carleton, T. *et al.* Global evidence for ultraviolet radiation decreasing COVID-19 growth rates. *PNAS* **118**(1), e2012370118 (2021).
23. Gorman, S., & Weller, R.B. Investigating the potential for ultraviolet light to modulate morbidity and mortality from COVID-19: A narrative review and update. *Front. Cardiovasc. Med.* **7**, 616527 (2020).
24. Merow, C. & Urban, M. C. Seasonality and uncertainty in global COVID-19 growth rates. *PNAS* **117**(44), 27456–27464 (2020).
25. Sagripanti, J. L. & Lytle, C. D. Inactivation of influenza virus by solar radiation. *Photochem. Photobiol.* **83**, 1278–1282 (2007).
26. Hemmes, J. H., Winkler, K. C. & Kool, S. M. Virus survival as a seasonal factor in influenza and poliomyelitis. *Nature* **188**, 430–431 (1960).
27. Nguyen, M. T., Silverman, A. I. & Nelson, K. L. Sunlight inactivation of MS2 coliphage in the absence of photosensitizers: Modeling the endogenous inactivation rate using a photoaction spectrum. *Environ. Sci. Technol.* **48**, 3891–3898 (2014).
28. Weberand, Th. P. & Stilianakis, N. I. A note on the inactivation of influenza A viruses by solar radiation, relative humidity and temperature. *Photochem. Photobiol.* **84**, 1601–1602 (2008).
29. de Arruda, E. *et al.* Acute respiratory viral infections in ambulatory children in urban northeast Brazil. *J. Infect. Dis.* **164**, 252–258 (1991).
30. Alonso, W. J. *et al.* Seasonality of influenza in Brazil: A traveling wave from the Amazon to the subtropics. *Am. J. Epidemiol.* **165**, 1434–1442 (2007).
31. Dbouk, T., & Drikakis, D. Weather impact on airborne coronavirus survival. *Phys. Fluids* **32**, 093312 (2020).
32. Ianevski, A. *et al.* Low temperature and low UV indexes correlated with peaks of influenza virus activity in northern Europe during 2010–2018. *Viruses* **11**, 207 (2019).
33. Gunthe, S.S. *et al.* On the global trends and spread of the COVID-19 outbreak: Preliminary assessment of the potential relation between location-specific temperature and UV index. *J. Public Health Theory Pract.* (2020).
34. Chen, S. *et al.* COVID-19 and climate: Global evidence from 117 countries. *medRxiv* preprint. https://doi.org/10.1101/2020.06.04.20121863.
35. Taagi, H. *et al.* The higher temperature and ultraviolet, the lower COVID-19 prevalence—Meta-regression of data from large US cities. *Am. J. Infect. Control* **48**, 1279–1285 (2020).
36. Isaia, G. *et al.* Does solar ultraviolet radiation play a role in COVID-19 infection and deaths? An environmental ecological study in Italy. *Sci. Total Environ.* https://doi.org/10.1016/j.scitotenv.2020.143757 (2020).
37. Walrand, S. Autumn COVID-19 surge dates in Europe correlated to latitudes, not to temperature-humidity, pointing to vitamin D as contributing factor. *Sci. Rep.* **11**, 1981 (2021).
38. Takagi, H. *et al.* Higher temperature, pressure, and ultraviolet are associated with less COVID-19 prevalence: Meta-regression of Japanese prefectural data. *Asia Pac. J. Public Health* **32**(8), 520–522 (2020).
39. Carvalho, F. R. S. *et al.* Potential of solar UV radiation for inactivation of coronaviridae family estimated from satellite data. *Photochem. Photobiol.* **97**(1), 213–220 (2020).
40. Biasin, M. *et al.* UV-A and UV-B can neutralize SARS-CoV-2 infectivity. *medRxiv* preprint. https://doi.org/10.1101/2021.05.28.21257989 (2021).
41. Zhu, N. *et al.* A novel coronavirus from patients with pneumonia in China 2019. *N. Engl. J. Med.* **382**, 727–733 (2020).
42. Cobey, S. Modeling infectious disease dynamics—The spread of the coronavirus SARS-CoV-2 has predictable features. *Science* **368**, 713 (2020).
43. Monto, A. S. *et al.* Coronavirus occurrence and transmission over 8 years in the HIVE cohort of households in Michigan. *J. Infect. Dis.* **222**, 9–16 (2020).
44. Li, Y., Wang, X. & Nair, H. Global seasonality of human seasonal coronaviruses: A clue for postpandemic circulating season of severe acute respiratory syndrome coronavirus 2?. *J. Infect. Dis.* https://doi.org/10.1093/infdis/jiaa436 (2020).
45. Baker, R. E., Yang, W. & Vecchi, G. A. Susceptible supply limits the role of climate in the early SARS-CoV-2 pandemic. *Science* https://doi.org/10.1126/science (2020).
46. O'Connor, C., Courtney, C. & Murphy, M. Shedding light on the myths of ultraviolet radiation in the COVID-19 pandemic. *Clin. Exp. Dermatol.* **46**(1), 187–188 (2021).
47. Cherrie, M. *et al.* Ultraviolet A radiation and COVID-19 deaths in the USA with replication studies in England and Italy. *Br. J. Dermatol.* https://doi.org/10.1111/bjd.20093 (2021).
48. Rendell, R., Khazova, M. & Higlett, M. O'Hagan, impact of high solar UV radiant exposures in spring 2020 on SARS-CoV-2 viral inactivation in the UK. *J. Photochem. Photobiol.* **97**(3), 542–548. https://doi.org/10.1111/php.13401 (2021).
49. De Natale, G. *et al.* The evolution of Covid-19 in Italy after the spring of 2020: An unpredicted summer respite followed by a second wave. *Int. J. Environ. Res Public Health* **17**, 8708 (2020).
50. Nelson, K. L. *et al.* Sunlight-mediated inactivation of health-relevant microorganisms in water: A review of mechanisms and modeling approaches. *Environ. Sci. Process Impacts* **20**, 1089–1122 (2018).
51. Zempila, M. M. *et al.* TEMIS UV product validation using NILU-UV ground-based measurements in Thessaloniki, Greece. *Atmos Chem. Phys.* **17**, 7157–7174 (2017).
52. Gueymard, C. Parameterized transmittance model for direct beam and circumsolar spectral irradiance. *Solar Energy* **71**(5), 325–346 (2001).
53. van Doremalen, N. *et al.* Aerosol and surface stability of SARS-CoV-2 as compared with SARS-CoV-1. *N. Engl. J. Med.* **382**, 1564–1567 (2020).
54. Ratnesar-Shumate, S. *et al.* Simulated sunlight rapidly inactivates SARS-CoV-2 on surfaces. *J. Infect. Dis.* **222**, 214–222 (2020).
55. Stadnytskyi, V., Anfinrud, Ph., Bax, C. E. & Bax, A. The airborne lifetime of small speech droplets and their potential importance in SARS-CoV-2 transmission. *PNAS* **117**(22), 11875–11877 (2020).
56. Chin, A. W. H. *et al.* Stability of SARS-CoV-2 in different environmental conditions. *Lancet Microbe* **1**(1), E10 (2020).
57. Sagripanti, J. L. & Lytle, C. D. Sensitivity to ultraviolet radiation of Lassa, vaccinia, and Ebola viruses dried on surfaces. *Arch. Virol.* **156**, 489–494 (2011).
58. Madronich S. The Atmosphere and UV-B Radiation at Ground Level, in *Environmental UV Photobiology* (Young, A. R., Moan, J., Björn, L. O., & Nultsch, W. eds.) 1–39 (Springer, 1993).

59. Verity, R. *et al.* Estimates of the severity of coronavirus disease 2019: A model-based analysis. *Lancet Infect. Dis.* **20**(6), 669–677 (2020).

### Acknowledgements
The work presented in this paper has been carried out in the context of the activities promoted by the Italian Government and in particular, by the Ministries of Health and of University and Research, against the COVID19 pandemic. Authors are grateful to former and current INAF's Presidents, Prof. N. D'Amico and M. Tavani, for the support and for a critical reading of the manuscript. Authors are also grateful to Prof. A. Musarò, Dr. S. Covino, Dr. M. Elvis, and Dr. F. Fiore for critically reading the papers and providing useful comments and suggestions.

### Author contributions
F.N. prepared Figs. 2a and 3, 4, S1 and S2. G.S. prepared Figs. 1 and 2b. All authors contributed equally to the discussion of the results and review of the manuscript.

### Competing interests
The authors declare no competing interests.

### Additional information
**Supplementary Information** The online version contains supplementary material available at https://doi.org/10.1038/s41598-021-94417-9.

**Correspondence** and requests for materials should be addressed to F.N.

**Reprints and permissions information** is available at www.nature.com/reprints.

**Publisher's note** Springer Nature remains neutral with regard to jurisdictional claims in published maps and institutional affiliations.

# Supplementary Information

# Solar UV–B/A Radiation is Highly Effective in Inactivating SARS-CoV-2

**Authors:**
Fabrizio Nicastro[1*], Giorgia Sironi[2], Elio Antonello[2], Andrea Bianco[2], Mara Biasin[3], John R. Brucato[4], Ilaria Ermolli[1], Giovanni Pareschi[2], Marta Salvati[5], Paolo Tozzi[4], Daria Trabattoni[3], Mario Clerici[6]

**Affiliations:**

1 Italian National Institute for Astrophysics (INAF) – Rome Astronomical Observatory, Rome, Italy
2 Italian National Institute for Astrophysics (INAF) – Brera Astronomical Observatory, Milano/Merate, Italy
3 Department of Biomedical and Clinical Sciences L. Sacco, University of Milano, Milano, Italy
4 Italian National Institute for Astrophysics (INAF) – Arcetri Astrophysical Observatory, Firenze, Italy
5 Regional Agency for Environmental Protection of Lombardia (ARPA Lombardia), Milano, Italy
6 Department of Pathophysiology and Transplantation, University of Milano and Don C. Gnocchi Foundation, IRCCS, Milano, Italy

\* E-mail: fabrizio.nicastro@inaf.it

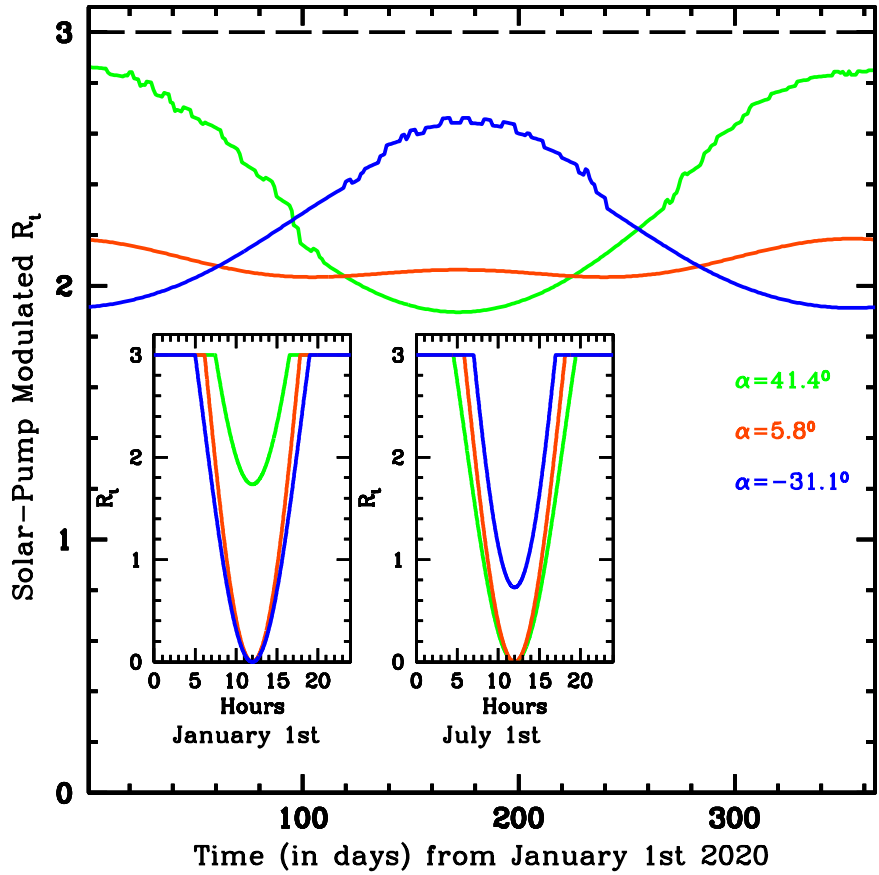

**Supplementary Figure S1 Main panel:** *day-by-day yearly modulation imprinted by the solar mechanism on the intrinsic reproduction number $R_0=3$ of the epidemic (black horizontal dashed line), at the three average latitudes of the countries of our N (green curves), T (orange curves) and S (blue curves) groups. Onsets: infra-day solar-pump modulation of $R_o$ at the three latitudes on January 1st (left onset) and July 1st (right onset) 2020, as labeled (see Methods for details).*

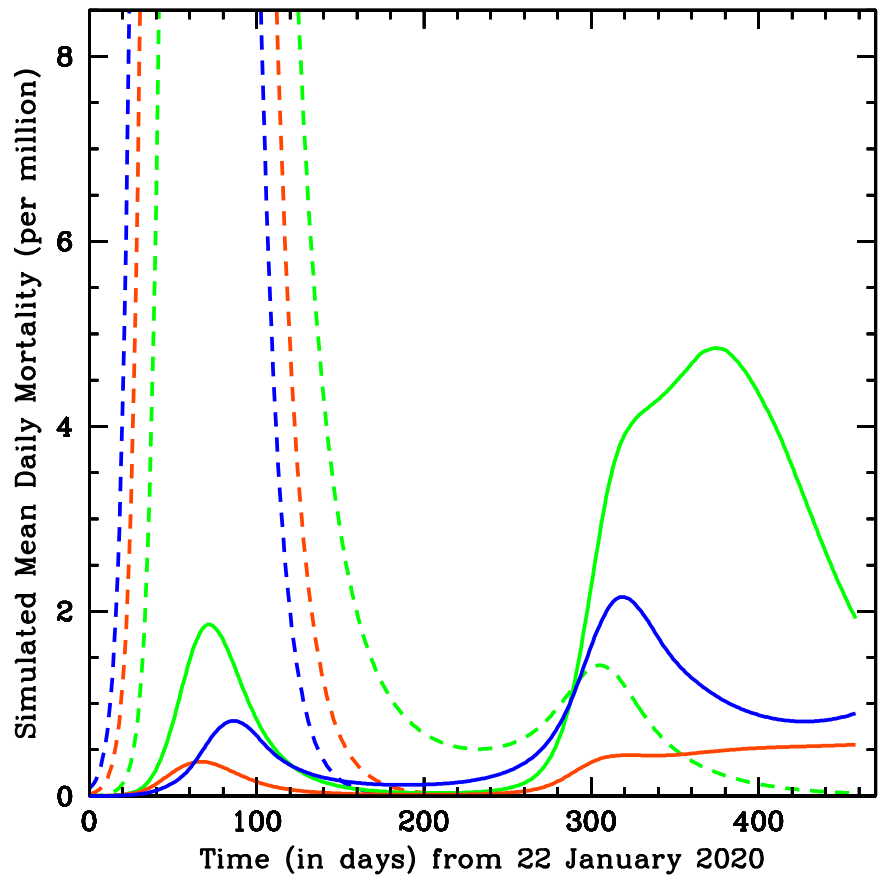

Supplementary Figure S2 Simulated mean daily mortality curves as a function of time, from 22 January 2020 through April 2021, *with (solid curves) and without (dashed curves) the effect of the Solar-Pump (see Methods for details).*